\documentclass[prl,a4paper,superscriptaddress,twocolumn,amsmath,amssymb]{revtex4-1}

\usepackage{graphicx}
\usepackage{color}
\usepackage{bm}
\usepackage{dsfont}
\usepackage{pdfpages}
\usepackage{hyperref}

\newcommand {\Fig}[1] {Figure~\ref{#1}}

\newcommand{\beq}{\begin{equation}}
\newcommand{\eeq}{\end{equation}}

\newcommand{\sithirty}{$^{30}$Si}
\newcommand{\sitwonine}{$^{29}$Si}
\newcommand{\sitwoeight}{$^{28}$Si}

\newcommand{\beqa}{\begin{eqnarray}}
\newcommand{\eeqa}{\end{eqnarray}}

\newcommand{\ket}[1]{\left| #1 \right\rangle}

\newcommand{\ttwo}{$T_2$}

\newcommand{\tonee}{$T_{\rm{1e}}$}

\newcommand{\ttwoe}{$T_{\rm{2e}}$}

\newcommand{\natsi}{$^{\rm nat}$Si}
\begin{document}

\title{Atomic clock transitions in silicon-based spin qubits}

\author{Gary Wolfowicz}
\email{gary.wolfowicz@materials.ox.ac.uk}
\affiliation{London Centre for Nanotechnology, University College London, London WC1H 0AH, UK}
\affiliation{Dept.\ of Materials, Oxford University, Oxford OX1 3PH, UK}  

\author{Alexei M. Tyryshkin}
\affiliation{Dept.\ of Electrical Engineering, Princeton University, Princeton, New Jersey 08544, USA}

\author{Richard E. George}
\affiliation{London Centre for Nanotechnology, University College London, London WC1H 0AH, UK} 

\author{Helge Riemann}
\author{Nikolai V. Abrosimov}
\affiliation{Institute for Crystal Growth, Max-Born Strasse 2, D-12489 Berlin, Germany}

\author{Peter Becker}
\affiliation{Physikalisch-Technische Bundesanstalt, D-38116 Braunschweig, Germany}

\author{Hans-Joachim Pohl}
\affiliation{Vitcon Projectconsult GmbH, 07745 Jena, Germany}

\author{Mike L. W. Thewalt}
\affiliation{Dept.\ of Physics, Simon Fraser University, Burnaby, British Columbia V5A 1S6, Canada}

\author{Stephen A. Lyon}
\affiliation{Dept.\ of Electrical Engineering, Princeton University, Princeton, New Jersey 08544, USA}

\author{John~J.~L.~Morton}
\email{jjl.morton@ucl.ac.uk}
\affiliation{London Centre for Nanotechnology, University College London, London WC1H 0AH, UK} 
\affiliation{Dept.\ of Electronic \& Electrical Engineering, University College London, London WC1E 7JE, UK} 

\date{\today}

\begin{abstract}
A major challenge in using spins in the solid state for quantum technologies is protecting them from sources of decoherence. This can be addressed, to varying degrees, by improving material purity or isotopic composition~\cite{Tyryshkin2011, Balasubramanian2009} for example, or active error correction methods such as dynamic decoupling~\cite{Viola1998, Bluhm2011}, or even combinations of the two~\cite{Steger2012, Maurer2012}.
However, a powerful method applied to trapped ions in the context of frequency standards and atomic clocks~\cite{Bollinger1985, Fisk1995}, is the use of particular spin transitions which are inherently robust to external perturbations. Here we show that such `clock transitions' (CTs) can be observed for electron spins in the solid state, in particular using bismuth donors in silicon~\cite{George2010, Morley2010}. This leads to dramatic enhancements in the electron spin coherence time, exceeding seconds. We find that electron spin qubits based on CTs become less sensitive to the local magnetic environment, including the presence of $^{29}$Si nuclear spins as found in natural silicon. We expect the use of such CTs will be of additional importance for donor spins in future devices~\cite{Pla2012}, mitigating the effects of magnetic or electric field noise arising from nearby interfaces.
\end{abstract}

\maketitle

Out of the various candidates for solid state qubits, spins have been of particular interest due to their relative robustness to decoherence compared to other degrees of freedom such as charge. So far, the most coherent solid state systems investigated have been the spins of well-isolated donors in bulk 28-silicon, with coherence times (\ttwo) of up to seconds (extrapolated) for the electron spin~\cite{Tyryshkin2011} and minutes for the nuclear spin~\cite{Steger2012}, comparable to those of ion trap qubits \cite{Haljan2005, Langer2005}. However, in practical devices, spin coherence times are likely to be limited by factors such as coupling to nearby qubits and magnetic or electric field noise from the environment. For example, cross-talk with other donors 100~nm away limits the electron spin \ttwoe~to a few milliseconds~\cite{Tyryshkin2011}, while a nearby interface can limit the donor electron spin \ttwoe~to 0.3~ms at 5.2~K \cite{Schenkel2006}. Finally, without isotopic enrichment, the 5$\%$ natural abundance of \sitwonine~limits the electron spin \ttwoe~to less than 1~ms~\cite{George2010, Morley2010}.

An approach to creating more robust qubits is to tune free parameters of the system Hamiltonian to obtain insensitivity to specific sources of decoherence. This has been extensively used in ion trap qubits to protect against magnetic field fluctuations \cite{Haljan2005, Langer2005}, building on work on atomic clocks where hyperfine states, used as frequency standards, must remain stable against such variations. These so-called ``clock transitions'' (CTs) have a transition frequency ($f$) which is insensitive to magnetic field ($B$) variations, at least to first-order (in other words $df/dB=0$). More recently, superconducting circuit qubits have also taken advantage of a tuned Hamiltonian to remain immune to charge, flux or current noise~\cite{Vion2002,Koch2007}. 

Nuclear spin CTs in rare-earth dopants (nuclear spins $I > 5/2$) have been studied in the context of optical quantum memories~\cite{Longdell2006, McAuslan2012} leading to a 600-fold improvement of the coherence times to 150~ms, limited by second-order effects, while recent experiments on phosphorus donor nuclear spins also exploited a CT~\cite{Steger2012}. For electron spins in the solid-state, CTs remain relatively unused due in part to the requirement of a spin Hamiltonian of sufficient complexity. One of the richest single-defect spin systems is the bismuth donor in silicon (Si:Bi), which possesses an electron spin $S=1/2$ coupled to a nuclear spin $I=9/2$. The electron spin decoherence rates for Si:Bi have been found to follow $df/dB$ in both natural silicon~\cite{George2010}, and isotopically enriched \sitwoeight~\cite{Wolfowicz2012}. These results, combined with the identification of a number of CTs in the spin Hamiltonian of Si:Bi~\cite{Mohammady2010, Mohammady2012}, motivate the study of spin coherence times around CTs in Si:Bi, where $df/dB\rightarrow0$. In this Letter, we investigate one such CT in Si:Bi, at 7.0317~GHz, using both natural silicon and \sitwoeight.

\begin{figure*}[t] 
\centerline{\includegraphics[width=\textwidth]{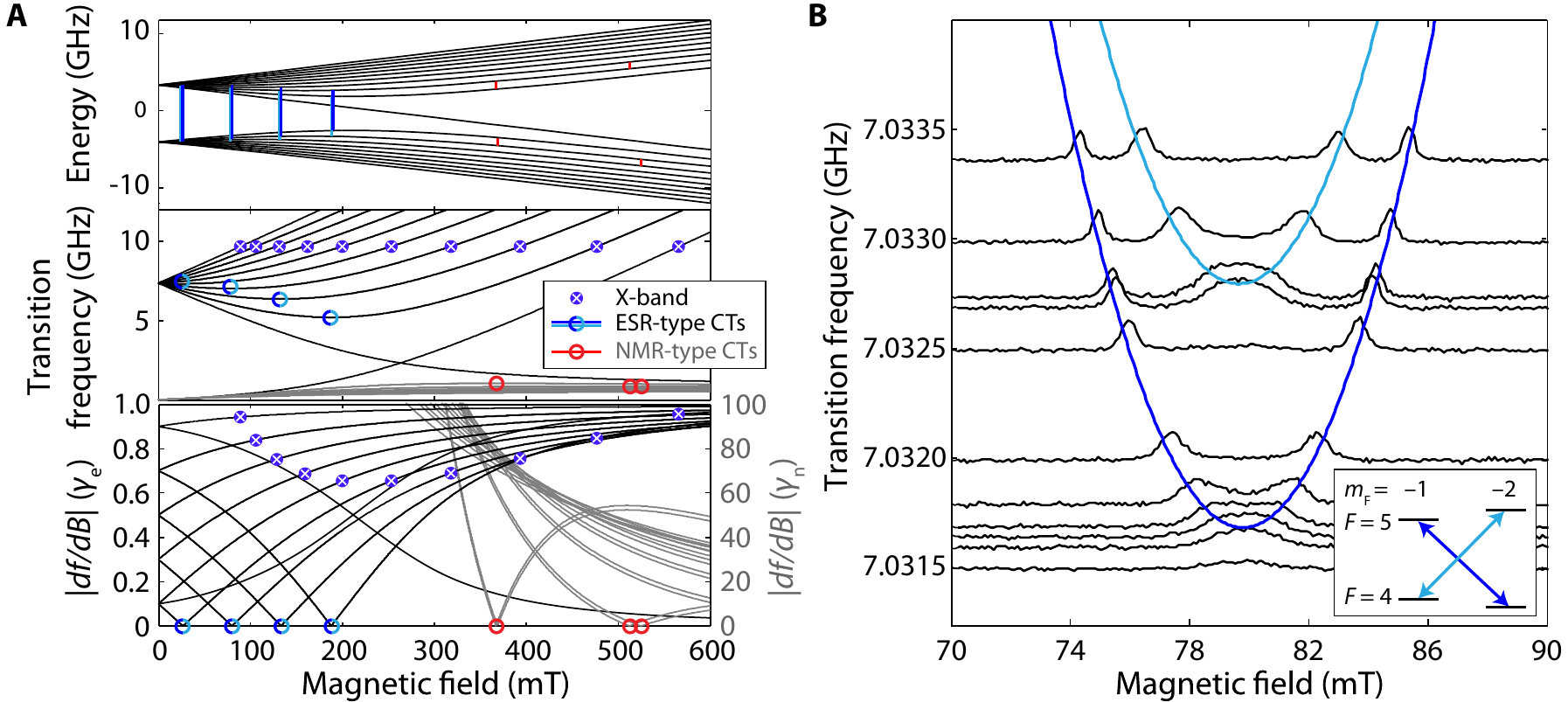}}\caption{\textbf{Electron spin resonance (ESR)-type clock transitions (CTs) of Si:Bi}. {\bf A,} The eigenstate energies (top) of Si:Bi as function of magnetic field, the ESR- (black) and NMR-type (grey) transition frequencies between these states (middle), and the first-order magnetic field dependence ($df/dB$) of these transition frequencies (bottom).
ESR-type CTs (blue lines and open circles) are found at 27, 80, 133 and 188~mT, and appear in the spectrum as doublets $\Delta F \Delta m_F = \pm1$ separated by up to 3~$\rm{MHz}$. NMR-type CTs are found above 300~mT (red lines and open circles). {\bf B,} Electron spin echo-detected magnetic field sweeps around the 80~mT CT measured at microwave frequencies $\geq 7.0315$~GHz. The transition probablities for $\Delta F\Delta m_F = +1$ (dark blue) and $-1$ (light blue) transitions are equal near the CT.} 
\label{fig:SiBiCT}
\end{figure*}

When describing the states of coupled electron and nuclear spins, two basis conventions are typically used: in the high magnetic field limit, the electron and nuclear spin projections $m_S$ and $m_I$ are good quantum numbers, while in the zero-field limit, the total spin $F$ ($= I \pm S$) and its projection $m_F$ ($=m_S+m_I$) are used. CTs are often found in an intermediate regime~\cite{Morley2012}, nevertheless it is possible to categorize them as nuclear magnetic resonance (NMR)- or electron spin resonance (ESR)-type, on the basis of whether the transition couples primarily to $S_x$ or $I_x$, where these are the electron and nuclear spin operators perpendicular to the applied magnetic field. The ESR-type CTs which we investigate in this manuscript involve states which are close to pure in the $|F,m_F\rangle$ basis and hence for convenience we label them according to the dominant $\ket{F,m_F}$ component (full details are given in the Supplementary Material and in Ref~\cite{Mohammady2010}).

For bismuth donors in silicon, NMR-type CTs can be found at high field ($>350$~mT) with frequencies around 1~GHz as shown in red in \Fig{fig:SiBiCT}A. At low field ($<200$~mT), four ESR-type CTs are present with frequencies in the range 5.2 to 7.3~GHz as shown in blue in the same figure. We will focus here on the ESR-type CTs, which possess only slightly reduced spin manipulation time compared to free electron spins as well as a large energy splitting even at low magnetic field (which has interesting applications for use in hybrid superconducting circuits \cite{George2010, Schuster2010, Kubo2012}). 


 

In the silicon samples we study here, Bi donors were introduced during crystal growth using the method developed in Ref~\cite{Riemann2006}, with concentrations ranging from $3.6\times10^{14}$~cm$^{-3}$ to $4.4\times10^{15}$~cm$^{-3}$. Pulsed-ESR experiments were performed using a spectrometer based around a modified Bruker Elexsys E580 system with a $\sim$7~GHz loop-gap cavity (for the CT) and 9.75~GHz dielectric resonator.


\Fig{fig:SiBiCT}B shows ESR spectra measured using microwave frequencies between 7.031 and 7.034~GHz, by plotting electron spin echo intensity as a function of magnetic field. 
%
The spectra show two transitions corresponding to $[\{\Delta F, \Delta m_F\} = \{\pm1,\pm1\}]$ and $[\{\Delta F, \Delta m_F\} = \{\pm1,\mp1\}]$; for brevity, these transitions can be distinguished by the value of the product $\Delta F\Delta m_F=\pm1$. Together, they offer a controllable two-qubit subsystem with low sensitivity to magnetic field fluctuations (see inset of \Fig{fig:SiBiCT}B). 

We model the ESR spectra using an isotropic spin Hamiltonian common for group V donors in silicon:
\begin{equation}
H_0 = B_0 (\gamma_e S_z \otimes \mathds{1} - \gamma_n \mathds{1} \otimes I_z) + A\vec{S}.\vec{I}
\label{eq:Hamiltonian}
\end{equation}
where the two first terms correspond to the electronic ($S$) and nuclear ($I$) spin Zeeman interactions with an external field $B_0$ and the last term corresponds to the hyperfine coupling $A$. A common way to estimate Hamiltonian parameters such as the electron and nuclear gyromagnetic ratios ($\gamma_e$ and $\gamma_n$) and the hyperfine constant is by measuring the magnetic field dependences of the spin transition frequencies. We use the opportunity provided by the CT (with $df/dB\rightarrow0)$ to extract a measure of the hyperfine constant $A = 1.47517(6)$~GHz with high precision, because uncertainties in the magnetic field become irrelevant. In our simulations, we additionally use the previously reported value of $\gamma_e = 27.997(1)$~GHz/T~\cite{Feher1959} and the generic value of $\gamma_n = 7$~MHz/T for $^{209}$Bi~\cite{George2010}.


\begin{figure}[t]%
\includegraphics[width=\columnwidth]{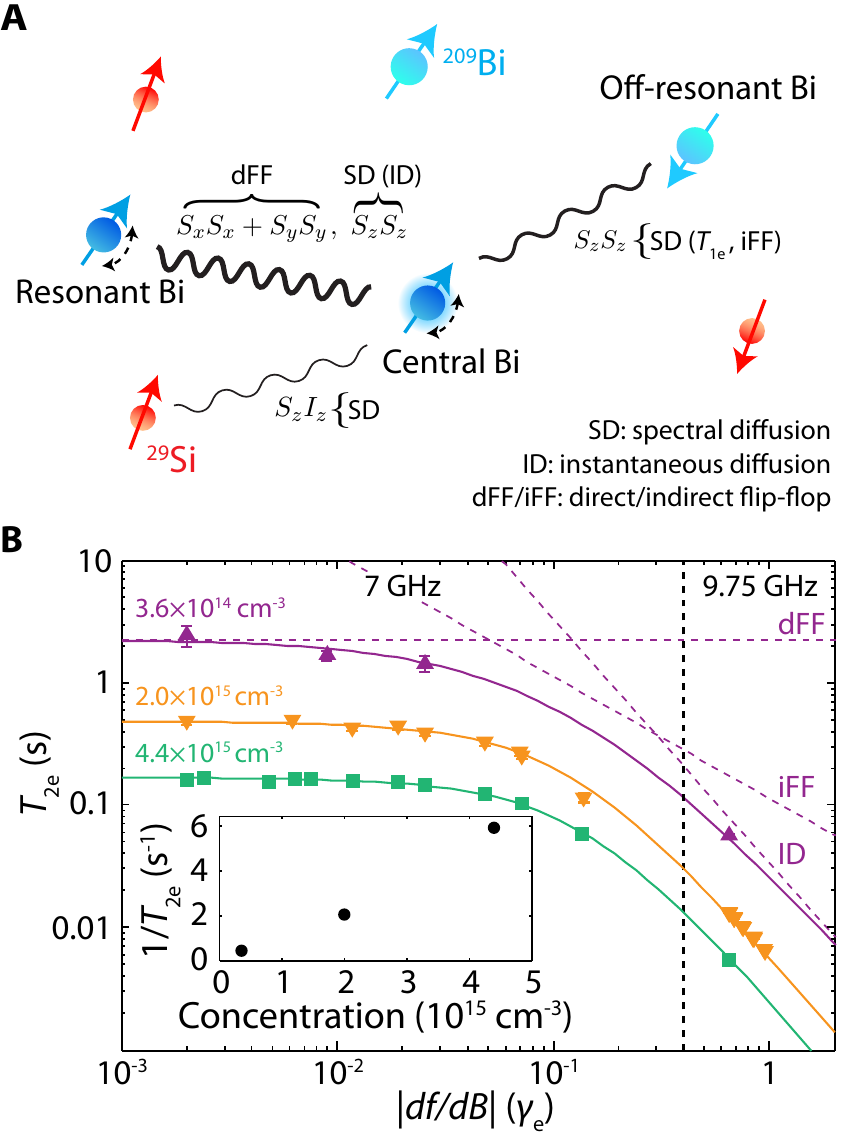}%
\caption{\textbf{Decoherence mechanisms of Bi donors in silicon and their dependence on $df/dB$.} \textbf{A,} In the central spin representation, a Bi donor is coupled to neighbouring Bi donors as well as \sitwonine\ spins. At the ESR CT, all spectral diffusion (SD) contributions to decoherence are essentially eliminated, leaving only the direct flip-flop term (dFF) between the central spin and a neighbouring, resonant Bi spin.
\textbf{B,} \ttwoe\ measurements at 4.8~K show a strong dependence on $df/dB$, as shown for 3 different donor concentrations in \sitwoeight:Bi. Measurements close to $df/dB = \gamma_e$ were taken using the ten X-band ESR transitions, while the remaining points were taken close to the CT. For each concentration, the dependence on $df/dB$ is modeled using contributions from ID, FF and iFF, as shown separately in dashed lines for the lowest concentration. 
Inset shows the limit of $1/$\ttwoe\ when approaching to the exact CT as a function of donor concentration, showing a nearly linear dependence, as expected for dFF.
}
\label{fig:DecoMeca}%
\end{figure}

\Fig{fig:SiBiCT}B shows that the ESR linewidth in the magnetic field domain increases around the CT: the derivative $df/dB$ tends to zero hence its inverse, $dB/df$, diverges until it becomes limited by the non-linear terms in $f(B_0)$. These spectra are all well fit assuming a constant linewidth in the frequency domain of 270~kHz. This linewidth can be attributed to a distribution in the hyperfine constant of around 60~kHz, using $\Delta f= \frac{df}{dA} \Delta A$ at the CT. Fourier-Transform ESR performed at a range of frequencies confirmed that the ESR linewidth in frequency domain is indeed magnetic field independent (see Supplementary Material).


We now examine the decoherence mechanisms which affect the electron spin of donors in silicon. At sufficiently low temperature ($< 5$~K), spin-lattice relaxation \tonee\ can be mostly neglected, and dipolar interactions ($\sim 2S_zS_z-(S_xS_x+S_yS_y)$) with neighbouring spins are the primary source of decoherence. In a central spin representation, as shown in \Fig{fig:DecoMeca}A, the surrounding spins can be divided into three categories: i) resonant spins affected by microwave excitation; ii) off-resonant spins of the same species, i.e. Bi spins in $m_F$ levels not addressed by the microwaves;  and iii) other spin species such as \sitwonine. Away from CTs, the limiting factor for electron spin coherence times is spectral diffusion (SD) from the $S_zS_z$ term of the dipolar interaction. This term can be assimilated into effective fluctuations in the magnetic field environment of the central spin. SD is independent of any frequency detuning between spins and thus is valid between the central spin and any others. 

In the static case, dipolar couplings to (ii) and (iii) can be refocused with a microwave $\pi$-pulse such as in the Hahn echo sequence. However, this does not correct for the dipolar coupling between resonant spins (i) as both spins are simultaneously flipped by the $\pi$-pulse. This is called ``instantaneous diffusion'' (ID) and limits \ttwoe\ to $\sim 10-100$~ms for typical donor concentrations ($>10^{14}$~cm$^{-3}$) \cite{Tyryshkin2011, Wolfowicz2012}~\footnote{By reducing the microwave power and effectively flipping only a small part of the resonant spins, it is possible to obtain an extrapolation of coherence times in the limit of no ID~\cite{Klauder1962,Salikhov1981}. However, it is not a solution to overcoming the effect of ID in practice.}. 
Furthermore, dynamic changes from spin flips in the environment cannot be refocused. At high temperature, such flips arise from phonon scattering but at low temperature, this is due to flip-flops (FF) from the $S_xS_x + S_yS_y$ term of the dipolar interaction. FF are energy conserving and as such are only relevant between spins that have similar transition frequencies. In natural silicon, the dominant decoherence mechanism is SD from \sitwonine\ FF, while in isotopically enriched \sitwoeight, it arises from FF between resonant Bi spin pairs. In the latter case, we distinguish between FF which involve the central spin (direct FF, dFF), and those which do not (indirect FF, iFF). 

We begin by discussing results on samples of isotopically enriched \sitwoeight\ (100~ppm \sitwonine). At the CT the transition frequency is insensitive to magnetic field fluctuations in first order, so we expect SD to have little effect, leaving only the dipolar coupling between resonant spin pairs. With reference to   \Fig{fig:DecoMeca}A, this implies then that all terms apart from dFF vanish.  
In \Fig{fig:DecoMeca}B, measurements of electron spin coherence times (\ttwoe) are shown for three different concentrations over a wide range of $df/dB$. The data includes values measured at X-band 
as well as those near the CT $(\Delta F\Delta m_F = +1)$ at 79.8~mT, 7.0317~GHz. Measurements at the CT shown here were taken at 4.8~K where $T_{\rm 1e} = 9$~s, however no increase in \ttwoe\ was seen at lower temperature. 

\begin{figure}[t]%
\includegraphics[width=\columnwidth]{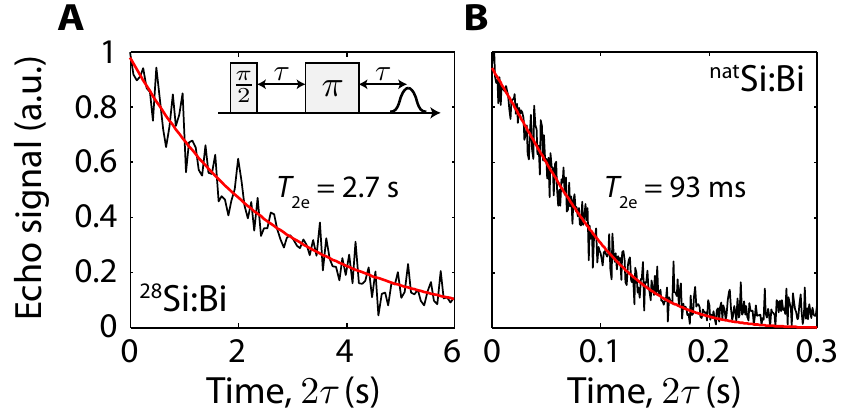}%
\caption{\textbf{Hahn echo decay at the CT.} 
\textbf{A,} \sitwoeight:Bi at 4.3~K with a Bi concentration of $3.6\times10^{14}$~cm$^{-3}$. 
\textbf{B,} \natsi:Bi at 4.8~K with a Bi concentration of $10^{15}$~cm$^{-3}$. The decay in natural Si is a stretched exponential, and therefore \ttwoe\ is defined as the time when the amplitude reaches 1/e. Magnitude detection was used to eliminate instrumental noise, likely due to phase noise in the microwave source.}%
\label{fig:T2eAll}%
\end{figure}

For each sample, enhancements of about two orders of magnitude are seen at the CT, compared to the case for a free electron g-factor, such as that of phosphorus donors. As shown in \Fig{fig:DecoMeca}B, the dependence of the measured \ttwoe\ on $df/dB$ arises from two factors: the effect on ID, and on iFF. 
ID has a known quadratic dependence on the gyromagnetic ratio of the central spin~\cite{Schweiger2001, Wolfowicz2012}, and becomes a negligible effect for $df/dB < 0.1 \gamma_e$.  Indirect FF dephase the central spin through the $S_zS_z$ term, giving a linear dependence of \ttwoe\ on $df/dB$. 
Direct FF, on the other hand, are not eliminated at the CT, and provide an upper bound on \ttwoe\ for a given donor spin concentration, as plotted in the inset of  \Fig{fig:DecoMeca}B. For the lowest concentration sample, electron spin coherence times of up to 2.7~s were measured from simple two-pulse Hahn echo decays, as shown in \Fig{fig:T2eAll}A. 

We now turn to measurements on Bi-doped natural silicon (\natsi:Bi), which has 5$\%$ \sitwonine. Away from the CT the effect of the \sitwonine\ ($I=1/2$) is both to broaden the ESR linewidth to about 0.4~mT (equivalent to 12~MHz in the frequency domain for a free electron) due to unresolved \sitwonine\ hyperfine, as well as to limit the \ttwoe\ to about 0.8~ms due to SD~\cite{George2010}. At the CT we find that the ESR linewidth reduces to 500~kHz (see Supplementary Material), within a factor of two of the value for enriched \sitwoeight\ material, while \ttwoe\ increases by over two orders of magnitude to about 90~ms (\Fig{fig:T2eAll}B).
The effect of the suppression of SD around the CT has been simulated for $^{\rm nat}$Si:Bi using cluster expansion methods~\cite{Balian2012}, though further refinements are required in the simulation before a quantitative comparison can be made. The stretched-exponential decay implies that \ttwoe\ is still limited at 93~ms by SD from \sitwonine\ due to the second order term $(d^2f/dB^2\neq0$). For modest \sitwoeight\ enrichment (e.g. [\sitwonine] $\approx$ 1000~ppm), \ttwoe\ should already exceed seconds, and indeed there may be an optimal \sitwoeight\ purity above which \ttwoe\ at the CT drops, due to the role of \sitwonine\ or \sithirty\ in detuning otherwise-identical spins~\cite{Witzel2010}.

We have shown how CTs in Si:Bi can be used to produce magnetic field-insensitive spin qubits with directly measured coherence times of several seconds. Such qubits would be insensitive to magnetic field noise arising, for example, from fluctuating dangling-bond spins at the Si/SiO$_2$ interface. Conversely, if electric field noise is dominant, this can couple to donor spins via the hyperfine interaction and cause decoherence. Again, CTs can be designed to be immune from electric charge noise by selecting points where $df/dA \rightarrow 0$ (see Supplementary Material). Through the use of CTs, it is likely that the seconds-long electron spin coherence times measured in the bulk can be harnessed for spins in practical quantum devices.

We thank Stephanie Simmons, Tania Monteiro and Setrak Balian for fruitful discussions. This research is supported by the EPSRC through the Materials World Network (EP/I035536/1) and a DTA, as well as by the European Research Council under the European Community's Seventh Framework Programme (FP7/2007-2013) / ERC grant agreement no. 279781. 
Work at Princeton was supported by NSF through Materials World Network (DMR-1107606) and through the Princeton MRSEC (DMR-0819860), and also by NSA/LPS through LBNL (6970579). J.J.L.M. is supported by the Royal Society. 

\bibliography{library}

\begin{thebibliography}{10}

\bibitem{Tyryshkin2011}
A.~M. Tyryshkin {\it et~al.}, Nature Materials {\bf 11},  143  (2012).

\bibitem{Balasubramanian2009}
G. Balasubramanian {\it et~al.}, Nature materials {\bf 8},  383  (2009).

\bibitem{Viola1998}
L. Viola and S. Lloyd, Physical Review A {\bf 58},  2733  (1998).

\bibitem{Bluhm2011}
H. Bluhm {\it et~al.}, Nature Physics {\bf 7},  109  (2011).

\bibitem{Steger2012}
M. Steger {\it et~al.}, Science (New York, N.Y.) {\bf 336},  1280  (2012).

\bibitem{Maurer2012}
P.~C. Maurer {\it et~al.}, Science (New York, N.Y.) {\bf 336},  1283  (2012).

\bibitem{Bollinger1985}
J. Bollinger, J. Prestage, W. Itano, and D. Wineland, Physical Review Letters
  {\bf 54},  1000  (1985).

\bibitem{Fisk1995}
P. Fisk {\it et~al.}, IEEE Transactions on Instrumentation and Measurement {\bf
  44},  113  (1995).

\bibitem{George2010}
R.~E. George {\it et~al.}, Phys. Rev. Lett. {\bf 105},  67601  (2010).

\bibitem{Morley2010}
G.~W. Morley {\it et~al.}, Nature Materials {\bf 9},  725  (2010).

\bibitem{Pla2012}
J.~J. Pla {\it et~al.}, Nature {\bf 489},  541  (2012).

\bibitem{Haljan2005}
P. Haljan {\it et~al.}, Physical Review A {\bf 72},  062316  (2005).

\bibitem{Langer2005}
C. Langer {\it et~al.}, Physical Review Letters {\bf 95},  060502  (2005).

\bibitem{Schenkel2006}
T. Schenkel {\it et~al.}, Applied Physics Letters {\bf 88},  112101  (2006).

\bibitem{Vion2002}
D. Vion {\it et~al.}, Science (New York, N.Y.) {\bf 296},  886  (2002).

\bibitem{Koch2007}
J. Koch {\it et~al.}, Physical Review A {\bf 76},  042319  (2007).

\bibitem{Longdell2006}
J. Longdell, A. Alexander, and M. Sellars, Physical Review B {\bf 74},  195101
  (2006).

\bibitem{McAuslan2012}
D. McAuslan, J. Bartholomew, M. Sellars, and J. Longdell, Physical Review A
  {\bf 85},  032339  (2012).

\bibitem{Wolfowicz2012}
G. Wolfowicz {\it et~al.}, Physical Review B {\bf 86},  245301  (2012).

\bibitem{Mohammady2010}
M. Mohammady, G.~W. Morley, and T.~S. Monteiro, Phys. Rev. Lett. {\bf 105},
  067602  (2010).

\bibitem{Mohammady2012}
M.~H. Mohammady, G.~W. Morley, A. Nazir, and T.~S. Monteiro, Phys. Rev. B {\bf
  85},  094404  (2012).

\bibitem{Morley2012}
G.~W. Morley {\it et~al.}, Nature Materials {\bf 11},  1  (2012).

\bibitem{Schuster2010}
D. Schuster {\it et~al.}, Phys. Rev. Lett. {\bf 105},  140501  (2010).

\bibitem{Kubo2012}
Y. Kubo {\it et~al.}, Phys. Rev. A {\bf 85},  012333  (2012).

\bibitem{Riemann2006}
H. Riemann, N. Abrosimov, and N. Noetzel, ECS Transactions {\bf 3},  53
  (2006).

\bibitem{Feher1959}
G. Feher, Phys. Rev. {\bf 114},  1219  (1959).

\bibitem{Note1}
By reducing the microwave power and effectively flipping only a small part of
  the resonant spins, it is possible to obtain an extrapolation of coherence
  times in the limit of no ID~\cite {Klauder1962,Salikhov1981}. However, it is
  not a solution to overcoming the effect of ID in practice.

\bibitem{Schweiger2001}
A. Schweiger and G. Jeschke, {\em {Principles of pulse electron paramagnetic
  resonance}} (Oxford University Press, Oxford, 2001).

\bibitem{Balian2012}
S.~J. Balian {\it et~al.}, Phys. Rev. B {\bf 86},  104428  (2012).

\bibitem{Witzel2010}
W. Witzel {\it et~al.}, Physical Review Letters {\bf 105},  187602  (2010).

\bibitem{Klauder1962}
J. Klauder and P. Anderson, Physical Review {\bf 125},  912  (1962).

\bibitem{Salikhov1981}
K. Salikhov, Journal of Magnetic Resonance (1969) {\bf 42},  255  (1981).

\end{thebibliography}

\newpage~\newpage
\includepdf[pages=1]{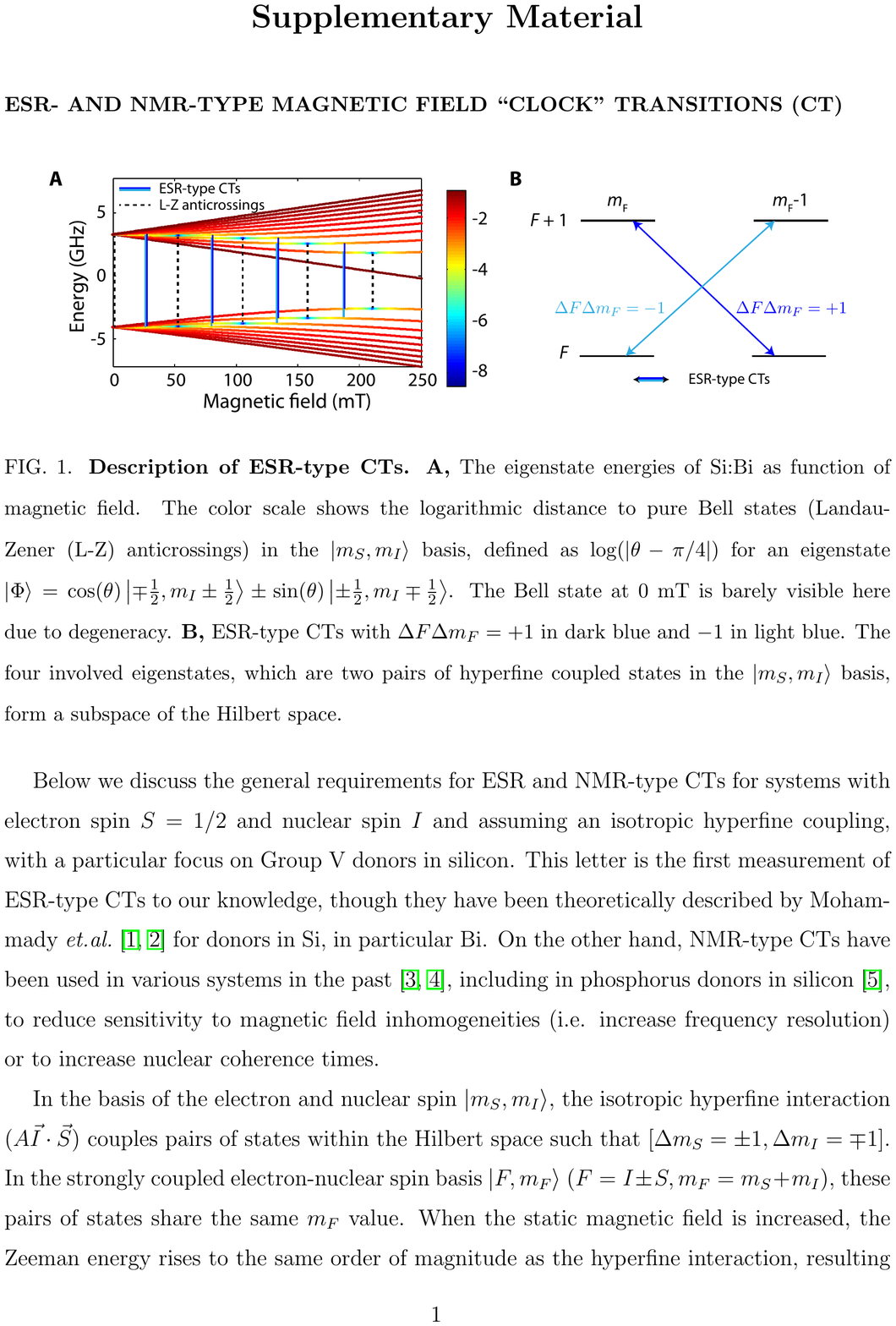}
\newpage~\newpage
\includepdf[pages=2]{SiBi-CT-SOM.pdf}
\newpage~\newpage
\includepdf[pages=3]{SiBi-CT-SOM.pdf}
\newpage~\newpage
\includepdf[pages=4]{SiBi-CT-SOM.pdf}
\newpage~\newpage
\includepdf[pages=5]{SiBi-CT-SOM.pdf}
\newpage~\newpage
\includepdf[pages=6]{SiBi-CT-SOM.pdf}

\end{document}